\documentclass[sigconf]{acmart}
\usepackage{graphicx}
\usepackage{enumitem}
\usepackage{xcolor}%
\usepackage{textcomp}%
\usepackage{tabularx}
\usepackage{comment}
\usepackage{graphicx}%
\usepackage{multirow}%
\usepackage{amsmath,amsfonts}%
\usepackage{amsthm}%
\usepackage{mathrsfs}%
\usepackage[title]{appendix}%

\usepackage{algpseudocode}%
\usepackage{listings}%

\usepackage{comment}
\definecolor{darkergreen}{RGB}{0,100,0} 

\usepackage{color, colortbl}
\usepackage{manyfoot}%
\usepackage{booktabs}%
\usepackage{algorithm}%
\usepackage{algorithmicx}%
\definecolor{shade}{gray}{0.95}
\newcommand{\mycomment}[1]{}

\AtBeginDocument{%
  }

\copyrightyear{2024}
\acmYear{2024}
\setcopyright{rightsretained}
\acmConference[L@S '24]{Proceedings of the Eleventh ACM Conference on Learning @ Scale}{July 18--20, 2024}{Atlanta, GA, USA} \acmBooktitle{Proceedings of the Eleventh ACM Conference on Learning @ Scale (L@S '24), July 18--20, 2024, Atlanta, GA, USA} \acmDOI{10.1145/3657604.3664694}
\acmISBN{979-8-4007-0633-2/24/07}

\begin{document}

\title{Towards Educator-Driven Tutor Authoring: Generative AI Approaches for Creating Intelligent Tutor Interfaces}


\author{Tommaso Cal{\`o}}
\orcid{0000-0002-3200-2348}
\affiliation{%
  \institution{Politecnico Di Torino}
  \streetaddress{Corso Duca degli Abruzzi, 24}
  \city{Torino}
  \state{Italy}
  \country{ITA}
  \postcode{10121}
}
\email{tommaso.calo@polito.it}

\author{Christopher J. MacLellan}
\affiliation{%
  \institution{Georgia Institute of Technology}
  \streetaddress{North Ave NW}
  \city{Atlanta}
  \state{Georgia}
  \country{USA}
  \postcode{30332}
}
\email{cmaclellan3@gatech.edu}


\begin{abstract}
Intelligent Tutoring Systems (ITSs) have shown great potential in delivering personalized and adaptive education, but their widespread adoption has been hindered by the need for specialized programming and design skills. Existing approaches overcome the programming limitations with no-code authoring through drag and drop, however they assume that educators possess the necessary skills to design effective and engaging tutor interfaces. To address this assumption we introduce generative AI capabilities to assist educators in creating tutor interfaces that meet their needs while adhering to design principles. Our approach leverages Large Language Models (LLMs) and prompt engineering to generate tutor layout and contents based on high-level requirements provided by educators as inputs. However, to allow them to actively participate in the design process, rather than relying entirely on AI-generated solutions, we allow generation both at the entire interface level and at the individual component level. The former provides educators with a complete interface that can be refined using direct manipulation, while the latter offers the ability to create specific elements to be added to the tutor interface. A small-scale comparison shows the potential of our approach to enhance the efficiency of tutor interface design. Moving forward, we raise critical questions for assisting educators with generative AI capabilities to create personalized, effective, and engaging tutors, ultimately enhancing their adoption.
\end{abstract}


\begin{teaserfigure}
  \includegraphics[width=\textwidth]{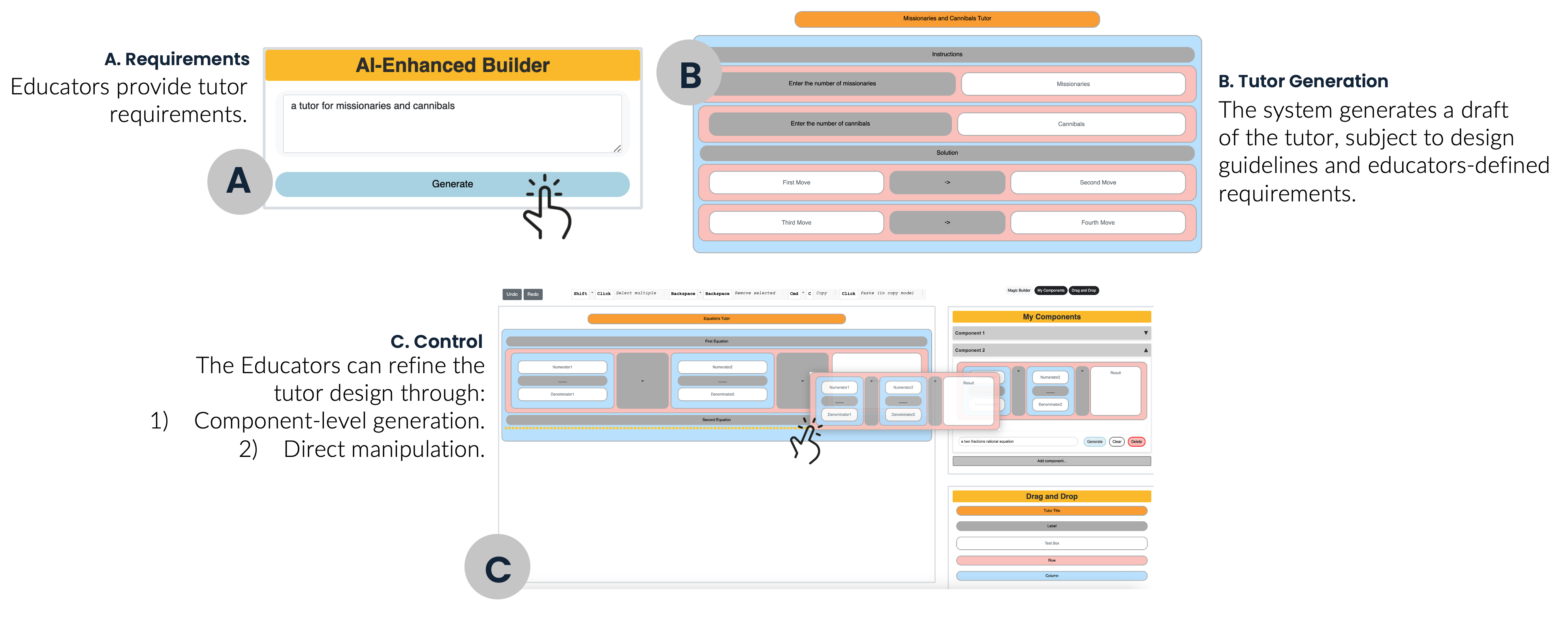}
\caption{Illustration of our AI-Enhanced Tutor Builder System: Educators provide input requirements (A) which inform the automated generation of a draft tutor interface (B), followed by the educator's hands-on refinement through component generation and direct manipulation (C), visualizing the integration of generative design and educator-driven customization.}
  \label{fig:teaser}
\end{teaserfigure}
\maketitle

\section{Introduction}

The rapid advancements in artificial intelligence is revolutionizing various sectors, and education is no exception. It is transforming the way we teach and learn, enabling personalized learning experiences that adapts to the unique needs and abilities of each student. Among the most promising applications of AI in education are Intelligent Tutoring Systems (ITSs), which have garnered significant attention from researchers and educators alike.

Intelligent Tutoring Systems (ITSs) provide personalized and adaptive education to students by supplying practice problems and sequences tailored to their expertise as well as support and feedback during problem solving. Research shows that ITSs are effective in raising learning outcomes \cite{ma2014intelligent}. By tracking each learner's knowledge and skills, ITSs can customize problem sequences to improve learning efficiency, allowing tutors to scale to serve many students simultaneously and address needs around providing supplemental instruction large classes~\cite{10.5555/1435351.1435353}.

However, several limitations prohibit the widespread use and applicability of ITSs. One significant limitation is that tutor development often requires specialized programming and tutor design knowledge~\cite{murray2003overview}. This prohibits non-technical educators from building their own customized tutors~\cite{ctat_tutor}, restricting creation to experts like scholars and software developers.

The recently introduced Apprentice Tutor Builder (ATB)~\cite{smith2024apprentice} enables educators to construct their own intelligent tutors. It proposes a drag-and-drop interface for assembling tutor layouts row and column formats, allowing them to interactively develop the underlying tutoring model with demonstrations through validating responses and supplying labels. While ATB mitigates the the challenges of tutor model authoring, \textit{it implicitly assumes educators can also effectively design interfaces without specific training or support}. In contrast, research shows that interface and instructional design requires specialized skills~\cite{Nielsen2023Bridging}.
An obvious question is therefore \textbf{how can we help educators in the UI design process to help them produce better tutors?}.

To address this, we propose to  enhance the Apprentice Tutor Builder with the capabilities of Generative AI~\cite{achiam2023gpt}. We create a prompt engineered to generate layouts and content directly from the requirements provided by the educators. This prompt incorporates design constraints to allow training of tutor, design principles, and examples to ensure the generated interfaces are effective and visually appealing. However, in order to avoid taking control away from educators in interface design, we enable generation of either the \textbf{whole interface} or just of \textbf{specific components}. The former serves as a starting point for the interface, and it can be further modified to meet educators' needs. The latter facilitates the creation of specific and reusable components and ensures educators for educators to place them within the final interface. With this dual approach, illustrated in Figure~\ref{fig:teaser}, we aim to balance the high automation provided by generative AI with high control for targeted customization. 
To our knowledge, this method represents a novel application in the field of tutor authoring; indeed, While AI has been leveraged to automate the creation of the underlying models in intelligent tutoring systems~\cite{MacLellan2016}, its potential to assist end-users, particularly educators, in designing the user interfaces of these systems has remained unexplored. Through a preliminary comparison, we show the potential of our method to accelerate the process of designing tutor interfaces, reducing the time and effort required to create both simple and complex interfaces. However, further research is needed to assess the quality and completeness of the AI-generated interfaces and their impact on the overall effectiveness of the tutoring system. We also raise questions about the optimal integration of generative interface tools into educators' workflows, ensuring the system meets real-world tutoring requirements. These questions will be addressed in a future user study, where we will further develop our approach. We aim to collect detailed feedback from educators on the usability and impact of the Generative AI assistance on their final design process and the resulting interfaces.
The final goal is to empower educators to create personalized and effective tutor interfaces by making the tutor design process more efficient and engaging with the assistance of generative AI. The resulting improvements in the usability of intelligent tutor interfaces could ultimately lead to wider adoption of adaptive education among students.

\begin{table*}[h!]
    \centering
    \renewcommand*\arraystretch{1.25}
    \small
    \begin{tabular}{l|ccc|ccc}
        \toprule
        \multirow{2}{*}{\textbf{Interface Type}} & \multicolumn{3}{c|}{\textbf{Time (s)}} & \multicolumn{3}{c}{\textbf{Keystrokes}} \\
        \cline{2-7}
         & \textbf{Classical} & \textbf{AI-Enhanced} & \textbf{Reduction} & \textbf{Classical} & \textbf{AI-Enhanced} & \textbf{Reduction} \\
        \midrule
        Simple & 187 & \textbf{143} & \textcolor{darkergreen}{\textbf{-23\%}} & 184 & \textbf{126} & \textcolor{darkergreen}{\textbf{-31\%}} \\
        \rowcolor{gray!20}
        Complex & 372 & \textbf{116} & \textcolor{darkergreen}{\textbf{-68\%}} & 141 & \textbf{74} & \textcolor{darkergreen}{\textbf{-47\%}} \\
        \bottomrule
    \end{tabular}
    \caption{Comparison of time and keystrokes required for building tutor interfaces: Classical vs. AI-Enhanced}
    \label{tab:interface_building_time_keystrokes}
\end{table*}

\subsection{Related Work}
The field of intelligent tutoring systems (ITSs) has witnessed significant advancements in recent years, with researchers and developers striving to create adaptive learning environments that cater to the unique needs of each student \cite{ma2014intelligent}. Despite the proven effectiveness of ITSs in enhancing learning outcomes, their widespread adoption has been hindered by the complexity involved in their development, which often requires specialized programming and design skills \cite{murray2003overview}. In an effort to democratize the creation of ITSs and empower educators to take an active role in their development, various authoring tools and platforms have emerged.

One prominent example is the Cognitive Tutor Authoring Tools (CTAT) \cite{ctat_tutor}, which provide a suite of tools for designing and deploying cognitive tutors. CTAT allows educators to create tutors by demonstrating problem-solving steps and specifying pedagogical rules. Similarly, the Authoring Software Platform for Intelligent Resources in Education (ASPIRE) \cite{aspire_tutor} enables domain experts to create constraint-based tutors by defining the domain model and problem-solving strategies. Another notable system is SimStudent \cite{matsuda2007simstudent, maclellan2014authoring} and the related Apprentice Learner System \cite{MacLellan2022}, which employ machine learning techniques to model student learning and support tutor authoring.

These tools primarily focus on the authoring of the domain model and pedagogical strategies, assuming that educators possess the necessary skills to design effective user interfaces. Our approach, in contrast, specifically addresses the challenges of assisting educators in creating tutor's user interfaces.

Generative user interfaces for end-user design have garnered significant attention in recent years, with various approaches being explored to automate or assist in the design process~\cite{Nielsen2023Generative}. One of the main line of research focuses on the unsupervised generation of user interfaces, where the system automatically creates interfaces based on a set of predefined rules or learned patterns. For example, Neural Design Network \cite{desnet} uses deep learning to generate user interface layouts from a given set of UI components. Closer to our work, Huang. et al. \cite{huang2021creating} introduced a method to generate layouts from textual descriptions using transformers models.

Another approach involves the controlled generation of sub-elements, where the system assists designers by generating specific components or suggestions based on user input. Stylette \cite{stylette} is a system that allows end-users to customize web elements based on natural language instructions, while Rewire \cite{rewire} suggests alternative layouts for a given user interface based on design heuristics and user feedback.

With respect to the above methods, our approach differs in two key aspects. First, it focuses on the application domain of tutor design. Second, it integrates both unsupervised generation at the interface level and controlled generation of components. This is in contrast to previous works, focused on the refinement of specific UI components \cite{stylette}, generation of layouts without human intervention \cite{desnet}, or layout generation without AI support \cite{articsle}.

\section{AI-Enhanced Tutor Interface Builder}
Our method for enhancing the ATB with generative AI capabilities involves three key components: a Domain Specific Language (DSL) for communicating with the LLM, prompt engineering to guide the generation process, and two levels of interaction for flexibility and control.

We define a compact DSL to represent tutor interface layouts, which includes fundamental elements such as \texttt{title[value]} for specifying the tutor's title, \texttt{row} and \texttt{column} for representing horizontal and vertical arrangements of elements, \texttt{label[value]} for defining text labels, and \texttt{input[placeholder]} for defining input fields with optional placeholder text. These elements can be combined to generate complex tutors. The DSL representation enables efficient communication with the LLM and ensures that the generated HTML output adheres to the desired template, preventing inconsistencies and deviations from the intended layout and aesthetics that may arise if a tutor's HTML interface is generated directly from the LLM.

To guide the generative model in creating appropriate tutor interfaces, we engineered a prompt consisting of different sections. The \textbf{System Description} provides an overview of the desired tutor interface, emphasizing the importance of a clear problem statement and a step-by-step resolution pathway to align with the educational objectives. The \textbf{Format Explanation} explains the DSL format used to represent the tutor interface layout, enabling the model to generate layouts that conform to the specified format. The \textbf{Design Instructions} specify design principles, such as separating input elements, arranging elements within rows and columns, and avoiding interactive buttons within the layout. The \textbf{Task Instruction} instructs the model to transform a detailed description of the tutor into the compact DSL representation, ensuring that the model understands its primary objective and generates the desired output format. Finally, a set of representative \textbf{Examples} serves as a reference for the model on how to apply the above principles in different contexts.

We introduce two levels of interface generation to fit into different design phases. First, \textbf{Interface Generation} supports creating a complete tutor interface layout based on the provided detailed description. It serves as a starting point for tutor designing. This allows the user to start from a generated user interface aligned with the design principles specified in the prompt.  We argue that the refined user interface will be more likely to follow these principles. In addition, interface generation could be particularly useful for users who prefer a more automated approach or have limited time for customization. On the other hand, \textbf{Component Generation} outputs designs for specific and reusable interface components; e.g., widgets for equations or other forms and flows. This capability aims to let users create interfaces that closely match their specific intent at a lower scope, while still benefiting from the efficiency and consistency provided by the generative model. 

Our AI-enhanced tutor interface builder is implemented on top of the existing ATB interface, leveraging an HTML/Javascript front-end and a Flask backend. We utilize GPT-4~\footnote{version gpt-4-0613} as the LLM engine to power the generation process. The interface and component generator are integrated into ATB's user interface as toolbar widgets.

\section{Preliminary Evaluation}

\begin{figure*}[ht]

\centering
\includegraphics[width=0.8\textwidth]{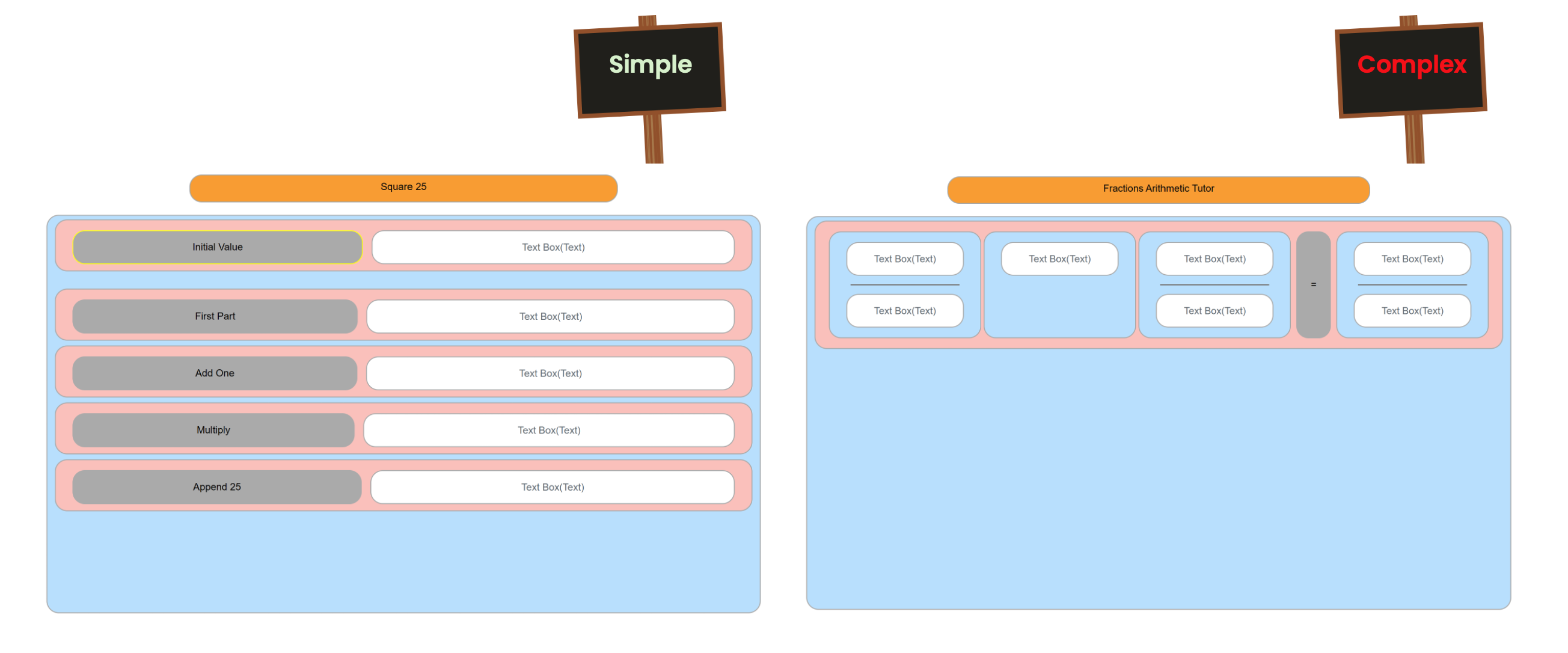}
\caption{This image illustrates the two interfaces used in the evaluation: On the left, the 'Simple' interface, designed for sequential problem-solving tasks, offers a user-friendly layout with simple input fields. On the right, the 'Complex' interface is tailored for an arithmetic equation solver, featuring a more advanced layout with multiple input fields and operational functions to handle equations.}
\label{fig:interface}
\end{figure*}

To evaluate the efficiency of the AI-enhanced Apprentice Tutor Interface Builder, we conducted a small-scale comparison with a previous version. We compared the performances of four team members against the reported performances of high-expertise individuals from the ATB paper. Although comparing our team's performance with that of the high-expertise ATB participants is not a strictly controlled comparison, we believe their high level of expertise significantly influences their performance, making it a suitable basis for preliminary comparison.  Furthermore, to support our findings and mitigate some of the issues associated with comparing different participant groups, we employed the Keystroke-Level Model~\cite{keystroke}. This model evaluates the efficiency of the interface by measuring the minimum number of keystrokes required to complete tasks, providing a quantitative measure of user interaction efficiency.

\paragraph{Evaluation Setup}
Users were asked to design the same two interfaces used in a prior ATB evaluation~\cite{smith2024apprentice}: a simple interface for a sequential problem and and a more complex interface for an arithmetic equation solver, both illustrated in Figure \ref{fig:interface}. We recorded the time taken for each task to compare the efficiency of the two approaches.

\paragraph{Results}
As shown in Table \ref{tab:interface_building_time_keystrokes}, using AI assistance consistently reduced the time required to build both simple and complex.
Interestingly, the efficiency gain was more pronounced in the case of the complex interface, with a 68\% reduction in time compared to the classical approach. This can be attributed to the fact that the complex interface, tailored for an arithmetic equation solver, required a more advanced layout with multiple input fields and operational functions. In this scenario, the AI assistance likely played a more significant role as users could leverage it to generate the equation components and layout elements more efficiently.
On the other hand, the efficiency gain for the simple interface was lower but substantial, at 23\%. This lower gain can be explained by the need for users to manually type all the labels in the simple interface, a process that was not necessary for the complex interface. This manual input likely offset some of the efficiency gains provided by the generative AI capabilities. Furthermore, although the reduction in absolute keystrokes is not as substantial as the time savings, it still supports the overall improvements in efficiency.

\paragraph{Limitations}
While this small-scale comparison provides promising indications of generative AI potential to improve tutor's interface design efficiency, particularly for complex tutors, further large-scale studies with diverse participants, tasks, and detailed feedback are necessary to validate these findings and gain deeper insights into the tool's impact on user satisfaction and the design process.

\section{Conclusions and Future Work}
In this paper, we presented an approach to improve the design of intelligent tutors' with Generative AI. The comparison demonstrates the potential of the approach to significantly increase efficiency of interface designing, particularly for complex interfaces.
To develop further our approach, we aim to address the following research questions:
\begin{itemize}
    \item \textbf{RQ1:} How can generative interface tools best integrate into educators' tutor building efforts?
    \item \textbf{RQ2:} What is the optimal configuration to balance design freedom and control for educators?
    \item \textbf{RQ3:} To what extent can automatic design tools meet the diverse requirements of real-world tutoring contexts?
\end{itemize}

To address these questions, we plan to conduct a study involving educators to assess our approach. This study will provide insights into how generative interface tools can best meet the needs of educators. Moreover, understanding how well these tools can accommodate the diverse requirements of real-world tutoring contexts is crucial for their successful adoption. By actively involving educators in the development process, we can identify interaction patterns that align our solutions with their needs, ultimately enhancing the quality and accessibility of intelligent tutoring systems.

Finally, we believe that our approach could be extended and applied to a wider range of tutor-building tools, such as CTAT~\cite{ctat_tutor}. By investigating the generalizability of our approach, we aim to unlock new ways of integrating generative AI in supporting educator-driven tutor design. Moreover, the concepts and methods presented in this paper could also find applications in other areas of end-user design, such as website builders or game development tools, opening up new possibilities for empowering non-expert users to create engaging and effective digital experiences.
\section{Acknowledgement}
This project is supported by National Science Foundation under Grant No. 2247790 and Grant No. 2112532. Any opinions, findings, and conclusions or recommendations expressed in this material are those of the author(s) and do not necessarily reflect the views of the National Science Foundation.
\bibliographystyle{ACM-Reference-Format}
\bibliography{DITS-LASLBR-2024-BIB}

\end{document}